\documentclass{SciPost}
\usepackage{booktabs} 
\hypersetup{
	colorlinks,
	linkcolor={red!50!black},
	citecolor={blue!50!black},
	urlcolor={blue!80!black}
}

\usepackage[bitstream-charter]{mathdesign}
\DeclareSymbolFont{usualmathcal}{OMS}{cmsy}{m}{n}
\DeclareSymbolFontAlphabet{\mathcal}{usualmathcal}

\fancypagestyle{SPstyle}{
	\fancyhf{}
	\lhead{\colorbox{scipostblue}{\bf \color{white} ~SciPost Physics }}
	\rhead{{\bf \color{scipostdeepblue} ~Submission }}
	
	\fancyfoot[C]{\textbf{\thepage}}
}

\usepackage{color}

\begin{document}
	
	\pagestyle{SPstyle}
	
	\begin{center}{\Large \textbf{\color{scipostdeepblue}{
	Geometric Prototype Learning in Quantum Hilbert Space with Matrix Product States\\
	}}}\end{center}
	
	\begin{center}\textbf{
			Kun Zhang\textsuperscript{1\S},
			Lei Ding\textsuperscript{2\S},
			Sheng-Chen Bai\textsuperscript{1},
			Jing Sun\textsuperscript{1},
			An-Qi Jing\textsuperscript{1},
			Min Tang\textsuperscript{2$\star$} and
			Shi-Ju Ran\textsuperscript{1$\dagger$}
	}\end{center}
	
	\begin{center}
		{\bf 1} Department of Physics, Capital Normal University, Beijing 100048, China
		\\
		{\bf 2} Department of Cardiac Arrhythmia, State Key Laboratory of Cardiovascular Disease, Fuwai Hospital, National Center for Cardiovascular Diseases, Chinese Academy of Medical Sciences and Peking Union Medical College, Beijing 100037, China
		\\
		\noindent\textsuperscript{\S}These authors contributed equally to this work.
		\\[\baselineskip]
		$\star$ \href{mailto:email1}{\small doctortangmin@yeah.net}\,,\quad
		$\dagger$ \href{mailto:email2}{\small sjran@cnu.edu.cn}
	\end{center}

	\section*{\color{scipostdeepblue}{Abstract}}
	\textbf{\boldmath{Quantum probability provides a novel framework for formulating machine-learning (ML) problems in Hilbert space. We introduce a prototype-based learning scheme where class representatives are encoded as generative matrix product states (MPS). Because these prototypes reside in the same Hilbert space as quantum-encoded data samples, various ML tasks such as classification and clustering can be performed through geometric measures of quantum states. This approach lifts prototype learning from classical feature space to quantum Hilbert space. Benchmarks on Fashion-MNIST and a real-world electrocardiogram dataset demonstrate that our method outperforms classical prototype approaches while remaining competitive with standard black-box neural networks. We also identify an ``attraction'' effect induced by the quantum-probabilistic prototypes and introduce a dimensionality-reduction scheme based on prototype distances. Our results establish quantum states as an explainable framework for prototype learning, opening new directions for designing ML algorithms in quantum Hilbert space.}}
	
	\vspace{\baselineskip}
	
	\noindent\textcolor{white!90!black}{
		\fbox{\parbox{0.975\linewidth}{
				\textcolor{white!40!black}{\begin{tabular}{lr}
						\begin{minipage}{0.6\textwidth}
							{\small Copyright attribution to authors. \newline
								This work is a submission to SciPost Physics. \newline
								License information to appear upon publication. \newline
								Publication information to appear upon publication.}
						\end{minipage} & \begin{minipage}{0.4\textwidth}
							{\small Received Date \newline Accepted Date \newline Published Date}
						\end{minipage}
				\end{tabular}}
		}}
	}

	\vspace{10pt}
	\noindent\rule{\textwidth}{1pt}
	\tableofcontents
	\noindent\rule{\textwidth}{1pt}
	\vspace{10pt}
	
	\section{Introduction}
	\label{sec:intro}
	
In modern machine learning (ML), achieving both high performance and model interpretability remains a fundamental challenge. Deep neural networks (NNs) have demonstrated remarkable success across diverse domains, from computer vision~\cite{krizhevsky2017imagenet} to biomedical signal analysis~\cite{esteva2017dermatologist}, yet their ``black-box'' nature often obscures the rationale behind predictions. This opacity poses significant obstacles in high-stakes applications. For example, in clinical diagnosis, attempting to generate post-hoc explanations for a black-box model can be dangerously misleading if the model secretly relies on spurious correlations or administrative artifacts rather than true physiological indicators~\cite{rudin2019stop}. These vulnerabilities have motivated a growing interest in inherently interpretable learning frameworks that combine strong predictive power with transparent, mathematically principled models.

Classical ML typically relies on high-dimensional feature spaces and complex nonlinear mappings, which, while expressive, often lack interpretability. Quantum machine learning (QML) offers a complementary perspective: by embedding data into Hilbert space, it provides a structured and probabilistically grounded environment for representation and inference~\cite{biamonte2017quantum}. Beyond potential computational advantages, such as quantum speedup~\cite{lloyd2014quantum, liu2021_quantum_speed_up}, the Hilbert-space framework allows learning algorithms to exploit intrinsic geometric and probabilistic structures, making similarity measure, classification, and clustering operations more transparent~\cite{NIPS2016_5314b967, havlivcek2019supervised_QuantumAdvantage, huang2022QuantumAdvantage}. Although the probabilistic advantages of QML have recently inspired preliminary applications in the medical domain—such as medical image analysis and clinical diagnostics~\cite{ur2022quantum}—these efforts remain relatively scarce and demand further systematic investigation. In this sense, QML is not only a tool for computational speedup but also a natural platform for developing interpretable learning schemes that operate directly on quantum states, especially for high-stakes fields like healthcare.

Tensor networks (TN), and in particular matrix product states (MPS), provide a powerful approach for efficiently representing exponentially large Hilbert spaces~\cite{verstraete2008matrix, SCHOLLWOCK201196, ran2020tensorbook}. Originally developed to describe strongly correlated quantum systems, TNs offer controlled, interpretable parametrizations of complex quantum states. More recently, TN methods have been applied to ML, demonstrating promise in both generative and discriminative tasks~\cite{PhysRevX.8.031012}. Despite these advances, most approaches treat TNs primarily as compact function approximators or probabilistic models, relying on classical post-processing or black-box readout layers to complete learning. The potential of TNs as native Hilbert-space learners, where both data and learnable representatives coexist and interact directly through intrinsic geometric relations, remains largely unexplored. This capability is particularly relevant for interpretable prototype-based learning, providing a natural bridge between quantum probabilistic representation and transparent decision-making.

In this work, we introduce quantum-probabilistic prototype learning (QPPL), a native Hilbert-space framework in which class prototypes are encoded as generative MPS's. Via a quantum feature map~\cite{NIPS2016_5314b967}, samples are mapped into quantum product states, which are defined in the same Hilbert space as that of the MPS prototypes. Tasks such as classification and clustering are performed directly in Hilbert space using geometric distances, without additional black-box readout layers. In this construction, the MPS prototypes act as quantum centroids, unifying representation, similarity measurement, and decision-making within a single formalism. We benchmark QPPL on Fashion-MNIST and a real-world electrocardiogram (ECG) dataset, showing it outperforms classical prototype methods while remaining competitive with widely used NNs. Furthermore, we identify an attraction effect induced by the MPS prototypes, which we leverage to design a secondary dimensionality-reduction scheme based on prototype distances, further improving clustering performance. These results establish QPPL as a transparent, mathematically principled framework where MPS prototypes do not merely summarize data, but actively reshape the data manifold as geometric attractors. By anchoring classification and clustering directly to quantum-probabilistic metrics, our work bridges tensor networks with interpretable learning, offering a rigorous foundation for future geometric algorithm design.
	
\section{Quantum-probabilistic prototypes with matrix product state}
\label{sec:Methods}
	
Consider a dataset consisting of $N$ samples $\boldsymbol{x}^{[n]}$ $(n=1,\ldots,N)$, each of which is characterized by $M$ features with $\boldsymbol{x}^{[n]} = (x_{1}^{[n]}, x_{2}^{[n]}, \ldots, x_{M}^{[n]})$. The first step is to map the samples into a quantum Hilbert space through a quantum feature map~\cite{NIPS2016_5314b967}. We regularize each feature as $x_{m}^{[n]} = (x_{m}^{[n]} - x_{\min})/(x_{\max} - x_{\min}) \in [0,1]$. Then we map each sample into a quantum product state of $M$ spins (dubbed as sample states) as	
	\begin{eqnarray}
		|{\varphi}(\boldsymbol{x}^{[n]})\rangle = \prod_{\otimes m=1}^M |v^{[n,m]}\rangle = \prod_{\otimes m=1}^M \sum_{s=1}^{d} 
		{v}^{[n,m]}_{s} |{s}_m\rangle,
		\label{eq-fm}
	\end{eqnarray}
where $\{\left| s_{m} \right\rangle\}$ ($s_{m} = 1,2,\ldots,d$) represent the computational basis of local dimension $d$ (with $d=2$ corresponding to qubits). $\boldsymbol{v}^{[n,m]} = \left[\cos\left(\frac{\theta\pi}{2}x^{[n]}_m\right), \sin\left(\frac{\theta\pi}{2}x^{[n]}_m\right) \right]^{T}$ parameterizes each spin rotation on the Bloch sphere, with hyperparameter $\theta$ controlling the maximal rotation.
	
A quantum-probabilistic prototype is modeled as generative MPS~\cite{PhysRevX.8.031012,PhysRevB.101.075135} as
\begin{eqnarray}
		|\Psi\rangle =  \sum_{s_{1} \ldots s_{M}=0}^{d-1} \sum_{a_{1} \ldots a_{M-1}=0}^{\chi-1} A^{[1]}_{s_{1} a_{1}} A^{[2]}_{s_{2} a_{1}a_{2}} \ldots  A^{[M-1]}_{s_{M-1} a_{M-2}a_{M-1}} A^{[M]}_{s_{M} a_{M-1}} \prod_{m=1}^M |{s}_m\rangle,
		\label{eq-MPS}
\end{eqnarray}	
where $\{s_m\}$ $(m = 1, \ldots, M)$ denote the physical bonds of dimension $d$ and $\{\alpha_{m}\}$ are the virtual bonds of dimension $\dim(\alpha_{m}) = \chi$ controlling the expressive power of the MPS's~\cite{verstraete2008matrix, ran2020tensorbook}. ${\boldsymbol{A}^{[m]}}$ are local tensors of MPS that are to be optimized.

Crucially, both the sample states and MPS prototypes reside in the same Hilbert space of dimension $d^M$, with the key difference being the entanglement structure. This enables a geometric ML framework based on well-defined distances between samples and prototypes. The choice of distance measure is essential. We use the negative logarithmic fidelity (NLF)
\begin{equation}
	D^{\text{NLF}}(|\psi\rangle, |\psi'\rangle)= -\ln \left| \langle \psi |\psi' \rangle \right|.
\end{equation}
 A distance of zero indicates equality up to a global phase. NLF avoids the exponential decay of standard fidelity in many-body systems\cite{orthogonalitycatastrophe}.

The MPS prototypes are trained to model the joint probability distribution of training data via maximizing state overlap. For generating a sample $\boldsymbol{y}$, the probability satisfies
\begin{equation}
	P\left(\boldsymbol{y}\right) = \left| \langle \varphi(\boldsymbol{y})|\mathrm{\Psi} \rangle \right|^{2}.
\end{equation}
The variational tensors ${\boldsymbol{A}^{[m]}}$ are optimized by minimizing the average NLF over the training set, analogous to minimizing the negative log-likelihood in classical generative models. Gradient descent updates are performed as $\boldsymbol{A}^{\left[m\right]}\gets{\boldsymbol{A}}^{\left[m\right]}-\eta\frac{\partial L}{\partial{\boldsymbol{A}}^{\left[m\right]}}$ with $L$ the average NLF and $\eta$ the learning rate.

At convergence, each generative MPS approximates the centroid of its assigned samples in Hilbert space, possessing approximately equal distances to the mapped sample states. For multi-class datasets, a separate MPS prototype can be assigned to each class. This allows the construction of a general geometric ML framework for tasks including classification, clustering, anomaly detection, and dimensionality reduction, all uniformly in Hilbert space.

\section{Hierarchical classification and anomaly detection}
\label{sec:RESULTS}

In QPPL, classification is naturally implemented through Hilbert-space geometry: a sample $\boldsymbol{y}$ is assigned to the class of the nearest quantum-probabilistic prototype (generative MPS). Denoting the prototype for the $k$th class as $|\Psi^{[k]}\rangle$, the predicted label is
\begin{equation}
\hat{k} = \arg\min_{k} D^{\text{NLF}}(|\Psi^{[k]}\rangle, |\varphi(\boldsymbol{y})\rangle),
\label{eq:classify}
\end{equation}
where $D^{\text{NLF}}$ is the negative logarithmic fidelity. This purely geometric decision rule requires no additional readout layer, making the classification process inherently transparent and interpretable.

\begin{table*}[tbp]
	\centering
	\footnotesize
	\setlength{\tabcolsep}{0pt}
	\caption{Comparison of Model across different classification tasks. Bold values indicate the best performance in each column.}
	\label{tab:combined_performance}
	\vspace{5pt}
	\begin{tabular*}{\textwidth}{@{\extracolsep{\fill}} l c c c c c c c c c c c }
		\toprule
		& \multicolumn{2}{c}{\textbf{SR/PVC}} & \multicolumn{3}{c}{\textbf{LVOT/RVOT}} & \multicolumn{3}{c}{\textbf{RCC/LCC}} & \multicolumn{3}{c}{\textbf{RC/AC/LC}} \\
		
		\cmidrule(lr){2-3} \cmidrule(lr){4-6} \cmidrule(lr){7-9} \cmidrule(lr){10-12}
		
		Model & Acc & Std & Acc & Std & AUROC & Acc & Std & AUROC & Acc & Std & AUROC \\
		\midrule
		FC   & \textbf{100\%} & \textbf{0} & 94.09\% & 0.0167 & 0.6044 & 93.75\% & 0.0442 & 0.6026 & 87.14\% & 0.0262 & 0.6342 \\
		CNN  & \textbf{100\%} & \textbf{0} & \textbf{95.00\%} & 0.0211 & 0.6786 & 96.88\% & 0.0125 & 0.6913 & 90.36\% & 0.0267 & 0.7685 \\
		LSTM & \textbf{100\%} & \textbf{0} & 93.41\% & 0.0167 & 0.6590 & 98.75\% & 0.0153 & 0.7516 & 87.14\% & 0.1244 & 0.7171 \\
		NC   & \textbf{100\%} & \textbf{0} & 76.82\% & 0.0310 & 0.8294 & 80.63\% & 0.0538 & 0.6653 & 60.71\% & 0.0723 & 0.9639 \\
		KNN  & \textbf{100\%} & \textbf{0} & 90.00\% & 0.0196 & 0.9134 & 96.25\% & 0.0234 & 0.8528 & 86.43\% & 0.0311 & 0.9870 \\
		
		
		GLVQ  & \textbf{100\%} & \textbf{0} & 77.73\% & 0.0255 & 0.8164 & 80.63\% & 0.0538 & 0.6530 & 62.50\% & 0.0687 & 0.9585 \\
		
		RBFN  & \textbf{100\%} & \textbf{0} & 76.59\% & 0.0154 & 0.8510 & 92.50\% & 0.0153 & 0.8658 & 65.00\% & 0.0290 & 0.9778 \\

		QPPL & \textbf{100\%} & \textbf{0} & 94.77\% & 0.0397 & \textbf{0.9138} & \textbf{99.38\%} & 0.0125 & \textbf{0.9077} & \textbf{92.14\%} & 0.0367 & \textbf{0.9888} \\
		\bottomrule
	\end{tabular*}
	
	\label{combined_performance}
\end{table*}

We evaluate QPPL on a real-world ECG dataset, considering four classification tasks of increasing complexity: detection of premature beats (SR/PVC) and their subsequent hierarchical anatomical localization (LVOT/RVOT, RCC/LCC, and RC/AC/LC). Unlike tabular datasets with structured features, ECG signals constitute time-series data, where explainable ML models typically lag behind NNs. According to clinical electrophysiology experts, these four tasks present varying levels of diagnostic difficulty. Distinguishing SR from PVC is the most straightforward, as PVCs exhibit highly distinct, wide QRS complexes. Clinically, premature ventricular contractions originating from the ventricular outflow tract (VOT) account for the majority of cases, often exceeding 90\% in reported cohorts. However, distinguishing whether VOT-origin PVCs arise from the left or right VOT remains challenging. This difficulty is further exacerbated when attempting fine-grained sublocalization within the RVOT (RC/AC/LC; right, left, and anterior cusps), where the extreme anatomical proximity of these cusps results in only subtle morphological differences across surface leads.

This clinical assessment aligns with the empirical results in Table~\ref{tab:combined_performance}: while almost all baseline models achieve near-perfect accuracy on the simplest SR/PVC task, the performance of classical models degrades significantly on the most challenging RC/AC/LC ternary task. QPPL achieves remarkable robustness, outperforming classical prototype methods, such as nearest-centroid (NC), k-nearest neighbors (KNN), generalized learning vector quantization (GLVQ), and radial basis function networks (RBFN) across all levels of complexity, particularly on the hardest task. Furthermore, it remains highly competitive with widely used NN architectures, including FCNN, CNN, and LSTM~\cite{6795963}. These results highlight the advantage of performing prototype learning directly in quantum Hilbert space, where a unified geometric representation enables accurate and interpretable classification even for capturing subtle physiological variations in complex temporal data.

Beyond classification, the same geometric structure naturally enables anomaly detection, building upon established deep learning surveys for time-series anomalies~\cite{TimeAD}, quantum-inspired detection frameworks~\cite{QuantumAD}, and the use of tensor networks to model one-class distributions in high-dimensional spaces~\cite{TNAnomoly}. While Ref.~\cite{TNAnomoly} excellently leverages tensor networks to parameterize classical decision boundaries for one-class distributions, it typically requires explicit regularizations to constrain the boundary. In contrast, our framework expands this frontier by preserving native quantum-probabilistic semantics, thereby enabling a regularization-free geometric inference scheme. Bypassing the need for black-box heuristics or specific quantum circuits, we formulate anomaly detection purely as a native geometric task in Hilbert space, utilizing the minimal NLF distance to generative MPS prototypes as a transparent and mathematically principled anomaly score. The distance between a sample and its nearest prototype provides a quantitative measure of deviation from the learned distribution, with larger distances indicating a higher likelihood of anomaly. We evaluate detection performance using the area under the receiver operating characteristic curve (AUROC)~\cite{hanley1982meaning}, which is a threshold-independent performance evaluation. From a probabilistic perspective, AUROC represents the probability that a randomly selected anomalous sample is assigned a higher anomaly score than a normal one~\cite{hanley1982meaning}. In our QPPL framework, the degree of anomaly for a given sample $\boldsymbol{y}$ is intrinsically quantified by its minimal NLF distance to the learned generative MPS prototypes, defining the geometric anomaly score as $S_{\text{QPPL}}(\boldsymbol{y}) = \min_{k} D^{\text{NLF}}(|\Psi^{[k]}\rangle, |\varphi(\boldsymbol{y})\rangle)$. Consequently, AUROC is evaluated as
\begin{equation}
	\text{AUROC}_{\text{QPPL}} = \frac{1}{N_{+} N_{-}} \sum_{i=1}^{N_{+}} \sum_{j=1}^{N_{-}} \mathbb{I}\left( S_{\text{QPPL}}(\boldsymbol{x}_{+}^{(i)}) > S_{\text{QPPL}}(\boldsymbol{x}_{-}^{(j)}) \right),
	\label{eq:AUROC_QPPL}
\end{equation}
where $N_{+}$ and $N_{-}$ denote the total number of anomalous and normal samples, respectively. The indicator function $\mathbb{I}$ evaluates to $1$ for $S_{\text{QPPL}}(\boldsymbol{x}_{+}^{(i)}) > S_{\text{QPPL}}(\boldsymbol{x}_{-}^{(j)})$, $0.5$ in the event of a tie, and $0$ otherwise.

In the clinical context of PVCs, the acute success rate of ablation varies substantially according to the site of origin. Ablation of VOT PVCs is associated with relatively high success rates, typically exceeding 80\%--90\%, whereas PVCs arising from other locations may have success rates as low as 50\%--60\%. Therefore, more accurate identification of VOT-origin PVCs is of practical importance. By improving the discrimination of VOT versus non-VOT origins, the proposed approach may assist clinicians in preprocedural planning and in estimating the likelihood of procedural success, thereby facilitating more informed decision-making and setting appropriate expectations for patients regarding treatment outcomes.

As summarized in Table~\ref{tab:combined_performance}, QPPL consistently achieves the highest AUROC across all tasks, outperforming both classical prototype methods and neural networks. This improvement can be attributed to the Hilbert-space representation, which captures subtle deviations in temporal patterns through global state overlaps, enhancing sensitivity to out-of-distribution samples without requiring additional scoring functions.

\begin{table}[htbp]
	\centering
	\footnotesize
	\caption{Comparative evaluation of GMPS and Deep Learning models for ECG classification. Bold values indicate the best performance in each column.}
	\label{tab:GMPS_DL_5Class_Hierachical}
	\vspace{5pt}
	\begin{tabular}{l c c c c}
		\toprule
		& \multicolumn{2}{c}{\textbf{5-class Classifier}} & \multicolumn{2}{c}{\textbf{Hierarchical Classifier}} \\
		\cmidrule(lr){2-3} \cmidrule(lr){4-5}
		Model & Acc & Std & Acc & Std \\
		\midrule
		FC   & 86.90\% & 0.0446 & 87.59\% & 0.0197 \\
		CNN  & 88.97\% & 0.0268 & 89.20\% & 0.0213 \\
		LSTM & 86.44\% & 0.0349 & 88.39\% & 0.0655 \\
		NC   & 56.67\% & 0.0427 & 53.11\% & 0.0247 \\
		KNN  & 83.33\% & 0.0186 & 83.56\% & 0.0178 \\
		GLVQ & 59.11\% & 0.0227 & 54.67\% & 0.0285 \\
		RBFN & 58.44\% & 0.0382 & 59.11\% & 0.0163 \\
		QPPL & \textbf{91.72\%} & 0.0245 & \textbf{93.33\%} & 0.0303 \\
		\bottomrule
	\end{tabular}
\end{table}

We further introduce hierarchical classification within QPPL that exploits task structure, aligning with established strategies in both deep learning architectures ~\cite{Texhiera} and expert-driven medical diagnostic schemes ~\cite{ECGhiera}. Our QPPL framework anchors the decision hierarchy directly to generative MPS prototypes in Hilbert space, enabling more accurate and interpretable multi-level inference. Separate sets of MPS prototypes are constructed for each level of the decision hierarchy. A sample is classified by sequentially traversing the hierarchy, selecting the nearest prototype at each level. For more details, please refer to the Appendix~\ref{app:ECG}. Compared to the multi-class model, this approach reduces class overlap at each stage and allows prototypes to focus on task-relevant features. As shown in Table~\ref{tab:GMPS_DL_5Class_Hierachical}, the hierarchical strategy improves classification accuracy for the majority of the evaluated ML models compared to the direct multi-class approach, with QPPL achieving the most significant performance and the highest overall accuracy. This demonstrates that the geometric organization in Hilbert space naturally supports the decomposition of complex tasks into successive decisions, enabling more accurate and interpretable multi-level inference.

Moreover, to assess the generality of QPPL across data modalities, we evaluate the framework on Fashion-MNIST~\cite{fashion}, a benchmark image dataset. Despite the high dimensionality and complex spatial correlations inherent in image data, QPPL outperforms classical prototype-based methods and only slightly underperforms compared to nearest-neighbor approaches. See details in the Appendix~\ref{app:fashion}. This provides further evidence that the Hilbert-space geometric representation generalizes across modalities, allowing prototypes to capture salient structures in both temporal and visual data. 

\section{Dimensionality reduction and clustering}

The geometric nature of QPPL further enables a natural formulation of dimensionality reduction and clustering directly in Hilbert space. To visualize the structure of the data, we employ t-distributed stochastic neighbor embedding (t-SNE)\cite{tSNE}, using pairwise distances defined either in the original Euclidean feature space or in Hilbert space via NLF.

\begin{figure*}[tbp]
	\centering
	\includegraphics[width=0.9\textwidth]{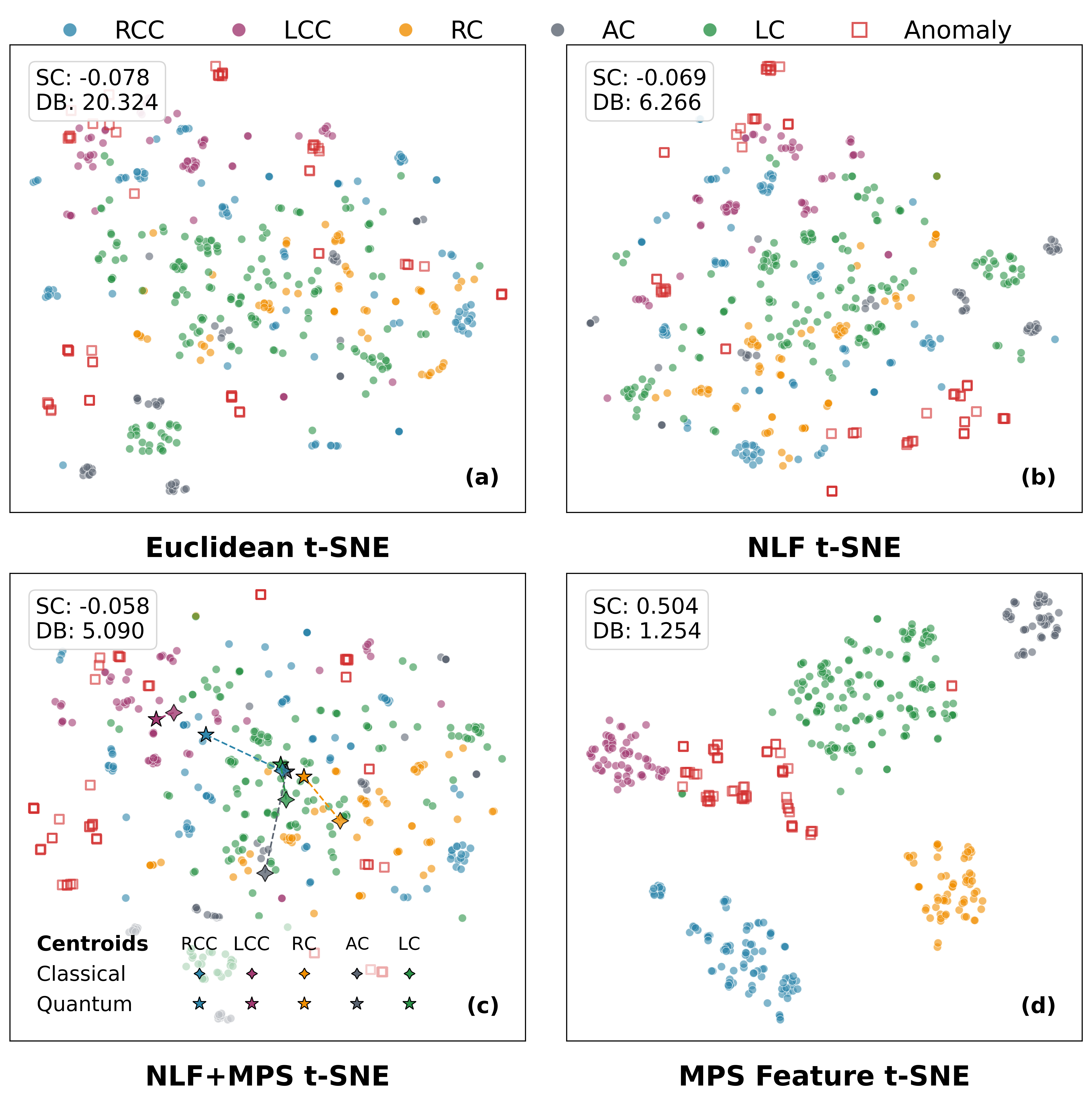}
	\caption{Comparative visualization of clustering structures. (a) Euclidean distance in the classical feature space. (b)–(c) NLF distance in the quantum Hilbert space. (d) MPS metric in the probability feature space.}
	\label{Fig.PCAtSNE}
\end{figure*}

A direct comparison reveals that the Hilbert-space geometry induced by QPPL leads to significantly improved low-dimensional representations. As shown in Fig.~\ref{Fig.PCAtSNE}(a,b), embeddings based on NLF distances of the ECG dataset exhibit clearer cluster separation and reduced overlap compared to those obtained using Euclidean distances $D_{n_1 n_2}^{\text{E}} = \left| \boldsymbol{x}^{[n_1]} - \boldsymbol{x}^{[n_2]} \right|$. This improvement is quantitatively supported by standard clustering metrics, including the Silhouette coefficient (SC)~\cite{ROUSSEEUW198753} and Davies–Bouldin index (DB)~\cite{4766909}. These results indicate that the quantum feature map, combined with fidelity-based geometry, provides a more discriminative representation of the data manifold.

Beyond this direct advantage, QPPL enables a further refinement by incorporating the learned MPS prototypes directly into the embedding process. Specifically, we construct an augmented distance matrix that includes not only pairwise distances between sample states, but also distances between prototypes and samples, as well as between prototypes themselves. This allows the prototypes to be explicitly embedded and visualized together with the data points in the same low-dimensional space.

Applying t-SNE to this augmented geometry yields a markedly enhanced clustering structure, as shown in Fig.~\ref{Fig.PCAtSNE}(c), with consistent improvements in both SC and DB metrics compared to embeddings based solely on sample distances. Notably, the embedded MPS prototypes do not coincide with the empirical centroids of the corresponding clusters in the low-dimensional space. Instead, they occupy distinct positions that reflect their role as generative representatives in Hilbert space, rather than simple averages of the embedded data points [see the dashed lines in Fig.~\ref{Fig.PCAtSNE}(c)].

Such a distinction is essential. While classical prototype methods typically define centroids directly in the feature space, the MPS prototypes are learned in the exponentially large Hilbert space and, generally speaking, are entangled states that encode global correlations and entanglement structure of the data distribution. As a result, their induced geometry cannot, in general, be reduced to Euclidean centroids. The observed discrepancy between prototype locations and cluster centroids thus provides an evidence that QPPL captures a nontrivial geometric organization of the data beyond classical representations. From this perspective, the improvement in clustering can be understood as arising from a prototype-induced reshaping of the distance landscape: the MPS prototypes act as geometric ``attractors'', reorganizing the embedding such that samples are drawn toward regions consistent with the learned quantum-probabilistic structure. 

In addition to distance-based embedding, QPPL naturally induces a more structured form of dimensionality reduction by directly leveraging the learned prototypes. Specifically, one can map each sample $\boldsymbol{x}^{[n]}$ to a vector of distances to all prototypes, as
\begin{equation}
	\boldsymbol{X}^{[n]} = \left[ D^{\text{NLF}}(|\Psi^{[1]}\rangle, |\varphi(\boldsymbol{x}^{[n]})\rangle), D^{\text{NLF}}(|\Psi^{[2]}\rangle, |\varphi(\boldsymbol{x}^{[n]})\rangle), \ldots \right],
	\label{MPS matric}
\end{equation}
where the dimensionality is given by the number of MPS prototypes. This construction defines a prototype-centered coordinate system, in which each axis quantifies the geometric proximity of a sample to a class-representative MPS in Hilbert space.

Such an embedding can be interpreted as a form of supervised metric learning in Hilbert space, where the learned prototypes define a data-adaptive metric that explicitly encodes class structure. Unlike conventional metric learning approaches that operate in Euclidean space, the QPPL framework leverages the quantum-probabilistic representation to capture global correlations and non-Euclidean geometry through fidelity-based distances. As a result, the induced representation integrates both local similarity (through pairwise distances) and global class-level information (through prototypes) within a unified geometric framework.

This perspective establishes explicit connections among dimensionality reduction, classification, and anomaly detection in QPPL. In all cases, inference is governed by distances to prototypes: nearest-prototype assignment yields classification, large distances signal anomalies, and the vector of prototype distances defines a compact representation. The resulting embedding $\{\boldsymbol{X}^{[n]}\}$ therefore preserves the discriminative structure learned in Hilbert space, leading to significantly improved clustering performance, as evidenced by enhanced SC and DB metrics [Fig.~\ref{Fig.PCAtSNE}(d)].


\section{Conclusion}
In this work, we introduced quantum-probabilistic prototype learning (QPPL), a native Hilbert-space framework in which both data and learnable representations are formulated as quantum states. By encoding class representatives as generative matrix product states (MPS), we established a unified geometric paradigm in which representation, similarity, and inference are intrinsically integrated through fidelity-based distances. Within this construction, prototype learning is lifted from classical feature space to quantum Hilbert space, enabling transparent and interpretable decision-making without auxiliary readout mechanisms.
	
We demonstrated that this geometric formulation provides a consistent foundation for a range of learning tasks. Classification arises from nearest-prototype assignment, anomaly detection from deviations in prototype distances, and dimensionality reduction from both pairwise geometry and prototype-induced embeddings. In particular, we identified a nontrivial attraction effect, whereby learned MPS prototypes actively reshape the geometric organization of data, enhancing clustering and separability. Furthermore, hierarchical classification emerges naturally from this framework, reflecting the ability of Hilbert-space geometry to decompose complex decision processes into successive, interpretable stages. Our results point toward a principled learning paradigm in which high performance and interpretability emerge not as competing objectives, but as natural consequences of a single underlying structure.



\section*{Acknowledgements}

\paragraph{Author contributions}
	K.Z., S.-C.B., J.S., and A.-Q.J. conducted the numerical simulations; S.-J.R. conceived the idea and designed the simulations; S.-J.R., L.D. and M.T. jointly contributed to improve and refine the core idea; L.D. and M.T. provided the medical data and relevant analyses. All authors contributed to the writing of the manuscript.
	
\paragraph{Funding information}
	 This work was supported in part by the Strategic Priority Research Program of Chinese Academy of Sciences (Grant No. XDB1270000) and CAS. The numerical simulations were partially performed on the robotic AI-Scientist platform of Chinese Academy of Sciences.
	

\begin{appendix}
\numberwithin{equation}{section}

\section{ECG dataset and preprocessing}
\label{app:ECG} 
	The ECG dataset comprises 86 standard 12-lead electrocardiogram (ECG) recordings. This includes 81 recordings exhibiting premature ventricular contractions (PVCs) of various anatomical origins and 5 recordings with normal sinus rhythm (SR). Among the PVC data, 26 cases originate from the left ventricular outflow tract (LVOT), and 55 cases originate from the right ventricular outflow tract (RVOT). The LVOT data consist of 14 cases from the right coronary cusp (RCC) and 12 from the left coronary cusp (LCC). The RVOT data comprise 8 cases from the anterior cusp (AC), 32 from the left cusp (LC), and 15 from the right cusp (RC).
	
	A single ECG recording contains multiple heartbeat waveforms. To expand the sample size for machine learning, the continuous ECG data required segmentation. In this study, to preserve the physically meaningful medical feature of the "interval between two heartbeats," each recording was segmented such that every sample contains exactly two consecutive heartbeats. Following this segmentation and MinMax normalization to the $[0, 1]$ interval, a total of 491 machine learning samples were generated. The sample distribution is as follows: RC (71), AC (48), LC (175), RCC (91), LCC (65), and SR (41).
	
	Furthermore, to explicitly evaluate the anomaly detection capability of the framework, specific out-of-distribution samples were selected for each task: 63 samples for the LVOT/RVOT binary classification, 36 samples for the RCC/LCC binary classification, and 27 samples for the RC/AC/LC ternary classification. For all classification and anomaly detection tasks, five independent experiments were performed, during which the machine learning samples were partitioned into training (80\%) and testing (20\%) sets using specific random seeds (42 to 46). See details in Table~\ref{tab:sample_distribution_horizontal}.
	
	\begin{table}[htbp]
		\centering
		\caption{Distribution of primary ML samples (left) and OOD anomaly samples (right) for the ECG dataset.}
		\label{tab:sample_distribution_horizontal}
		\begin{tabular}{lc @{\hspace{3em}} lc}
			\toprule
			\multicolumn{2}{c}{\textbf{Primary ML Samples}} & \multicolumn{2}{c}{\textbf{OOD Anomaly Samples}} \\
			\cmidrule(r){1-2} \cmidrule(l){3-4}
			Category & Number & Task & Number \\
			\midrule
			RC  & 71  & LVOT/RVOT Binary & 63 \\
			AC  & 48  & RCC/LCC Binary   & 36 \\
			LC  & 175 & RC/AC/LC Ternary & 27 \\
			RCC & 91  &                  &    \\
			LCC & 65  &                  &    \\
			SR  & 41  &                  &    \\
			\midrule
			\textbf{Total} & \textbf{491} &  &    \\
			\bottomrule
		\end{tabular}
	\end{table}
	
	Since each sample in the ECG dataset comprises 12 electrode signal channels, a dedicated GMPS state is trained for each individual channel during the training process. Consequently, each class is characterized by a set of 12 distinct GMPS states. In the classification task, the proximity of a sample $\boldsymbol{y}$ to class $k$ is evaluated by averaging NLF across all 12 electrodes. The predicted category for sample $\boldsymbol{y}$ is then determined by substituting this averaged metric into Eq. \ref{eq:classify}. Furthermore, in the visualization of the clustering results (as shown in panel (c) of Fig. \ref{Fig.PCAtSNE}), the positions of the quantum centroids were determined following the same channel-averaging methodology.
	
	The hierarchical decision process is structured as a top-down tree to classify the anatomical origins of premature ventricular contractions (PVCs). For a given PVC sample, the root-level classifier first determines its broad anatomical region, discriminating between the left ventricular outflow tract (LVOT) and the right ventricular outflow tract (RVOT). Based on this initial routing, localized branch classifiers are deployed in the second stage: a binary classifier distinguishes between RCC and LCC for samples identified as LVOT, while a ternary classifier differentiates among RC, AC, and LC for those identified as RVOT. At each node of this hierarchy, the decision is transparently governed by the minimal NLF distance to the corresponding local MPS prototypes.

\section{Fashion-MNIST Dataset}
\label{app:fashion}
Fashion-MNIST\cite{fashion} consists of $28 \times 28$ grayscale images of 10 categories of fashion products (e.g., T-shirts, trousers, and coats). Compared to the ECG dataset, Fashion-MNIST presents higher-dimensional feature spaces ($M=784$ after flattening) and complex spatial correlations that are typical of computer vision tasks.
	 
	 \begin{table}[htbp]
	 	\centering
	 	\footnotesize
	 	\caption{Comparative evaluation of GMPS and Deep Learning models for Fashion-MNIST.}
	 	\label{tab:fashion}
	 	\vspace{5pt}
	 	\begin{tabular}{l c c c c}
	 		\toprule
	 		\cmidrule(lr){2-5}
	 		& \multicolumn{2}{c}{Training Set} & \multicolumn{2}{c}{Testing Set} \\
	 		\cmidrule(lr){2-3} \cmidrule(lr){4-5}
	 		Model & Acc & Std & Acc & Std \\
	 		\midrule
	 		FC   & 98.21\% & 0.0011 & 87.74\% & 0.0055 \\
	 		CNN  & 99.49\% & 0.0013 & 90.42\% & 0.0053 \\
	 		LSTM & 92.80\% & 0.0284 & 86.00\% & 0.0024 \\
	 		NC   & 68.32\% & 0.0005 & 68.18\% & 0.0025 \\
	 		KNN  & 88.07\% & 0.0014 & 83.05\% & 0.0028 \\
	 		GLVQ & 75.01\% & 0.0005 & 74.48\% & 0.0013 \\
	 		RBFN & 82.78\% & 0.0005 & 82.23\% & 0.0029 \\
	 		GMPS & 95.62\% & 0.0401 & 84.79\% & 0.0427 \\
	 		\bottomrule
	 	\end{tabular}
	 \end{table}

In our experimental setup, the pixel intensities were normalized to the $[0, 1]$ interval and mapped into the quantum Hilbert space using the quantum feature map defined in the main text. A generative matrix product state (GMPS) prototype was trained for each of the 10 categories. The classification performance of our QPPL framework (denoted as GMPS) is compared against classical prototype methods and NN machine learning, with the results summarized in Table~\ref{tab:fashion}.

\end{appendix}


	
	
	
	
	\bibliography{newbib}
	

\end{document}